\documentclass[runningheads]{llncs}
\usepackage{amsmath,amssymb,amsfonts}
\usepackage{algorithmic}
\usepackage{graphicx}
\usepackage{textcomp}
\usepackage{listings}
\usepackage{xcolor}
\begin{document}
\title{Latent Semantic Structure in Malicious Programs}
%
%
\author{John Musgrave\inst{1}\and
Temesguen Messay-Kebede\inst{2}\and\\
David Kapp\inst{2}\and
Anca Ralescu\inst{1}}
\authorrunning{Musgrave J., et al.}
%
\institute{University of Cincinnati, Cincinnati, OH\\ \email{musgrajw@mail.uc.edu, ralescal@ucmail.uc.edu}
\and
Air Force Research Lab, Wright-Patt Air Force Base, Dayton, OH\\
\email{\{temesgen.kebede.1, david.kapp\}@us.af.mil}}
\maketitle              
\begin{abstract}
Latent Semantic Analysis is a method of matrix decomposition used for discovering topics and topic weights in natural language documents.  This study uses Latent Semantic Analysis to analyze the composition binaries of malicious programs.  The semantic representation of the term frequency vector representation yields a set of topics, each topic being a composition of terms.  The vectors and topics were evaluated quantitatively using a spatial representation.  This semantic analysis provides a more abstract representation of the program derived from its term frequency analysis.  We use a metric space to represent a program as a collection of vectors, and a distance metric to evaluate their similarity within a topic.  The segmentation of the vectors in this dataset provides increased resolution into the program structure.

\end{abstract}
\section{Introduction}
Analysis methods for malicious programs face challenges to the interpretation of results. This is due to the loss of semantic information when languages are translated between abstraction layers. A "semantic gap" exists, which is a gap between higher level language specification or theorem proving and the lower level instructions obtained from reverse engineering binary programs into their instruction sets. Proofs about the operational semantics of higher level languages are not easily translated to other layers in the abstraction hierarchy, such as binary or hex representations \cite{turi1997towards}, \cite{sebesta1999programming}.

\begin{figure}[h]
    \begin{lstlisting}
                    mov    ecx, rbp - 44
                    mov    eax, ecx
                    and    eax, 400
                    or     eax, 140
                    or     ecx, 1
                    cmp    rip + 170, 0
                    cmovne ecx, eax
                    mov    rbp - 44, ecx
                    mov    rip + 180, 0
                    jmp    0x100000000
    \end{lstlisting}
    \caption{Basic block segment of assembly instructions}
\end{figure}

A representation of operational semantics without abstraction cannot easily bridge the gap between architectural layers.  If program semantics are to be correlated across architectural layers, a greater level of abstraction is required.  Once a more abstract representation is found, then it is possible for programs to be classified by the identification of patterns in their structural properties.

Abstraction enables the potential identification of a process generating patterns of syntax to be uncovered.  Greater levels of abstraction allow for the ability to predict features of syntax.  These syntactic features can be predicted based on the structure of syntax provided by the abstraction.  In the case of Griffiths, Steyvers, and Tenenbaum, this was accomplished by the addition of a bi-partite graph with nodes representing concepts and words \cite{griffiths2007topics}, \cite{steyvers2005large}.

The strength of a semantic representation depends on the ability of the representation to express structural properties, to express syntactic patterns at a sufficient level of abstraction, and the degree of accuracy.  For greater accuracy we use a segmentation approach, which provides a more fine grained resolution.

One approach to providing both abstraction and structure is to discover the latent or hidden structure of a program by the analysis of terms in the document, called $tf-idf$ methods.  One advantage this offers is a representation not based on specific patterns of syntax.  A term frequency representation of a program contains latent structure with a greater level of abstraction, and datasets using term frequency representations can be analyzed with natural language analysis methods.  The abstraction of a latent structure would be a more descriptive feature of program semantic structure, as the correlation of syntactic features would be easily explained by the structure, and would allow for prediction.

Latent Semantic Analysis is already a widely used method of analysis for the discovery of topics in documents of natural language, and uses matrix decomposition to derive a more abstract representation with respect to a document's semantic content \cite{manning1999foundations}, \cite{shawe2004kernel}, \cite{griffiths2007topics}, \cite{steyvers2005large}.

We use Latent Semantic Analysis to provide a greater level of abstraction on a segmented term-frequency dataset.  We construct a metric space to evaluate the structure of a program through a spatial representation.  In order to have a measure of correlation in the metric space, we use Cosine Similarity as a similarity metric.

\subsection{Background}
Malware analysis has used Machine Learning methods to offer new solutions to program analysis.  Patterns of both syntax and semantics have been successfully classified using machine learning methods, but semantic program representation is still a difficult problem \cite{souri2018state}.

Latent Semantic Analysis has shown to be an effective method of analyzing natural language \cite{dumais2004latent}, \cite{griffiths2007topics}.

The process of Singular Value Decomposition (SVD) obtained from a linear algebraic approach allows the decomposition of matrices into orthonormal bases, along with weights \cite{manning1999foundations}, \cite{shawe2004kernel}, \cite{axler1997linear}.

\subsection{Related Work}
Machine Learning has been applied in many contexts to successfully identify malicious programs based on a variety of features.  Several different classification methods have been used for supervised learning including artificial neural networks and support vector machines.  Many datasets have been collected with several different kinds of features, including assembly instructions, n-gram sequences of instructions and system calls, and program metadata, \cite{souri2018state}, \cite{rawashdeh2021single}, \cite{kebede2017classification}, \cite{djaneye2019static}, \cite{chandrasekaran2020context}.

A number of studies have explored the use of static features at the level of file format, and their impact on the recognition of malicious programs. A decision tree was used as a method of classification for Windows PE files.  Subsequent studies have focused on malware classification using ensemble methods which include random forest with support vector machines and principal component analysis that was focused on file header features of Trojans. \cite{shafiq2009pe}, \cite{siddiqui2008detecting}, \cite{witten1999weka}.

While many studies of supervised learning rely on on $tf-idf$ representations for labeled data, n-gram sequences are typically computed linearly based on the placement of terms in the document as a whole, and are not segmented into blocks.  The instruction sequence is determined by the structural properties of the control flow graph, and not the placement in the document as in natural language documents.  This structure is often evaluated in isolation, and not included in the term frequency representation.  The evaluation of term frequencies in the document without segmentation assumes a linear sequence of terms with respect to their spatial placement, and this is not the case for executable binaries.  While edges in a control flow graph define the sequence order, sequential statements in the program are represented by nodes in the control flow graph.  This leads us to use a representation that is segmented in order to provide a suitable level of resolution.

The absence of accurate segmentation of a term frequency representation in program analysis would introduce noise from the existence of n-gram sequences that are not representative of deterministic instruction sequences.

\subsection{Outline}
Section 2 covers the experiments performed.  Section 3 covers the Results and Discussion.  Section 4 is a Summary and Conclusion.

\section{Experiment}
This section details the data collection and matrix decomposition methods that were performed.

\subsection{Data Collection}
An executable program is a sequence of instructions, each instruction being composed of opcodes and data operands. The dataset used in this study was composed by assembling a series of program binaries. Each binary program was reverse engineered using Linux decompilation tools such as GNU objdump to obtain an assembly representation of opcodes and data operands. This assembly representation was separated into segments called basic blocks, which are blocks of sequential instructions separated by a jump instruction. Figure 1 shows a basic block segment of contiguous assembly instructions from a binary program that was obtained after the decompilation process.  From a program segment we can obtain a sequence of terms and their term frequencies \cite{hopcroft2001introduction}, \cite{nar2019analysis}, \cite{hennessy2011computer}. 

Additional structure was obtain from the  program control flow graph in an adjacency matrix format was recovered using a concolic testing and symbolic execution tool $radare2$ for static analysis.

The following section outlines the data collection method and the data obtained.

\subsubsection{Define term dictionary}
An initial term dictionary was composed by selecting all possible opcodes present in the x86/64 opcode instruction set architecture. If we were to plot vectors in the space defined by all potential terms, each vector in the dataset would have a dimensionality of $527$ \cite{opcoderef}.

\subsubsection{Stemmed dictionary}
The term dictionary was then simplified by grouping terms that expressed the same arithmetic and logic or CPU control operation in terms of their operational behavior.  These terms may differ slightly in operating on different data types, e.g. floating point operations.  After grouping like terms in the dictionary the dimensionality of each vector in the term frequency vector space was reduced to $32$.  This eliminates unnecessary redundancy, and reduces the dimensionality of the metric space being considered for each vector \cite{opcoderef}, \cite{manning1999foundations}, \cite{shawe2004kernel}.

\subsubsection{Reverse engineering binary}
A set of binaries able to be executed within the x86/64 instruction set architecture were collected.  Each binary was analyzed by using the $objdump$ tool to recover the assembly code representations of the opcodes present in the binary executable.

\subsubsection{Convert program document into segments}
Each program sample was split into its component basic block segments by segmenting on each jump instruction or similar control state transition.  A jump instruction represents a transition of program control flow.

A basic block is a representation of a deterministic portion of the program. Each basic block was then represented as a vector by counting the frequencies of terms in the dictionary.

\subsubsection{Term distribution}
The distribution of opcode term frequencies is positively skewed towards data movement was measured.  The distribution is positively skewed and follows a power law distribution.

\subsection{Dataset}
The dataset resulting from the data collection contains two representations.  The first is a matrix vectors with one dimension per term in the term dictionary, and the numeric value representing the term frequency.  Each vector corresponds to a basic block segment of the program.  From a perspective of $tf-idf$, the program's basic blocks would be considered to be documents.  The document corpus is a collection of programs which comprise the dataset.

The second is an adjacency matrix representation of the program structure in terms of it's control flow graph edges. This captures the inter-segment structural composition in a network.

Figure 1 shows a basic block program segment of assembly instructions from a binary program that was obtained after the decompilation process. \cite{hopcroft2001introduction}, \cite{hennessy2011computer}.

\begin{figure}[t]
    \centering
    \[
    \begin{pmatrix}
    109.85 & 0 & 0 & 0 & 0 \\
    0 & 35.10 & 0 & 0 & 0 \\
    0 & 0 & 21.34 & 0 & 0 \\
    0 & 0 & 0 & 19.55 & 0 \\
    0 & 0 & 0 & 0 & 15.35
    \end{pmatrix}
    \]
    \caption{Top 5 weights of dimensions obtained from Singular Value Decomposition of term-document co-occurrence matrix.}
\end{figure}

\begin{figure*}[t]
    \centering
    \hspace{-2cm}
    \includegraphics[width=1.157\textwidth]{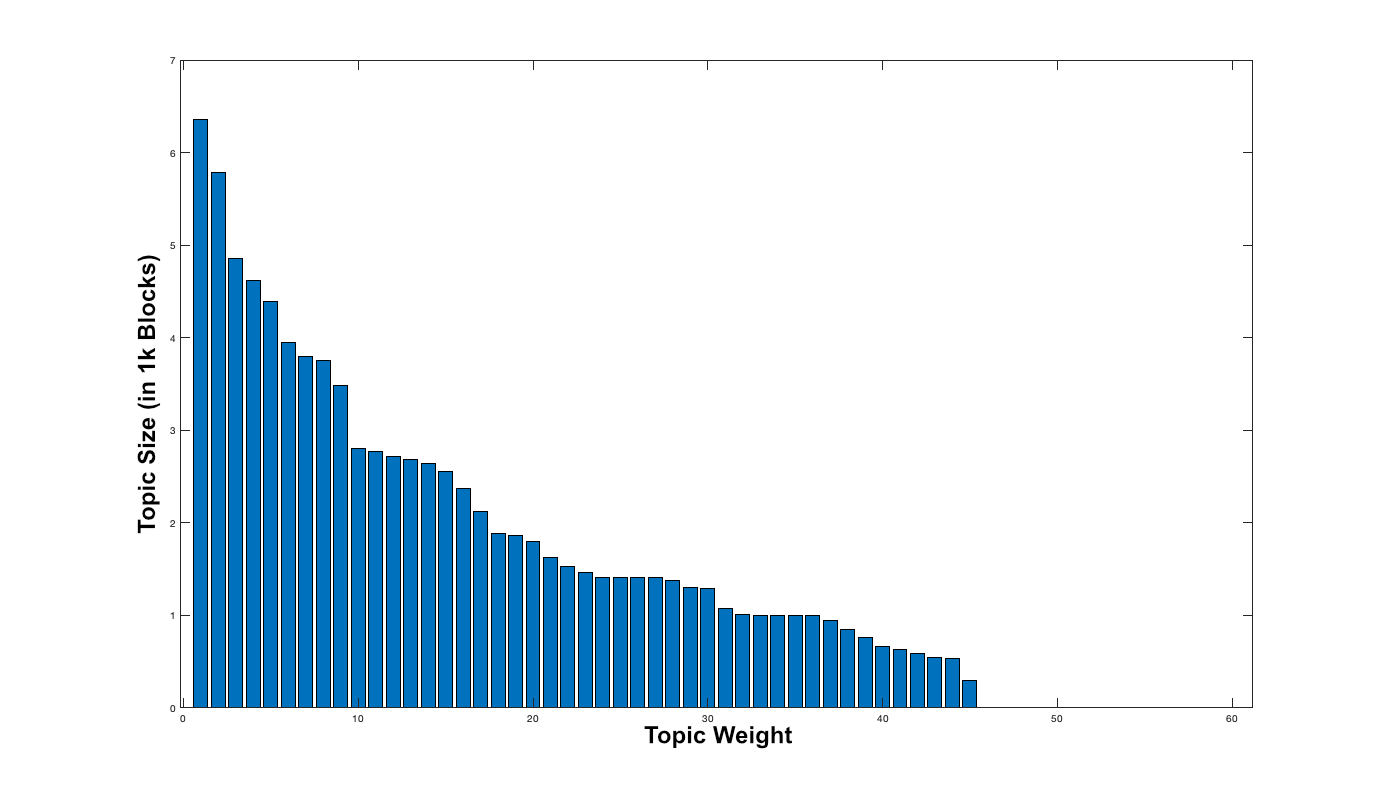}
    \caption{The distribution of discovered topic representations obtained from matrix decomposition of a program's control flow graph, which defines the structural relations between documents in the program.}
\end{figure*}

\subsection{Matrix Decomposition}
Singular Value Decomposition (SVD) was then performed on each of these datasets in order to decompose the matrices into their component parts.  Figure 2 shows the top weights of topic dimensions obtained from Singular Value Decomposition of term-document co-occurrence matrix.  These weights correspond to topic abstractions which are composed of multiple terms.  The control flow graph represents the structural relationships of the segments of the document, and for this reason this adjacency matrix was also decomposed into component matrices. In natural language documents, the syntactic structure is a linear sequence, but the sequence of program instructions is instead determined by program control flow edges in the adjacency matrix. For the matrices representing control flow networks, the SVD was performed on the adjacency matrix representation.

\subsubsection{Singular Value Decomposition}
The matrix of term frequencies decomposes into:

$U$ is an orthonormal basis for instructions in a segment.

$D$ is a matrix of weights per dimension (topic).

$V$ is an orthonormal basis for programs.

The othonormal basis for instructions in a segment shows the relationship between a document and its abstraction. The topic weight shows the strength of the individual abstract topic representation. The orthonormal basis for the program shows the degree to which a program is associated with an abstract representation or topic.  These components were evaluated using a spatial representation.

\subsubsection{Construction of Metric Space}
Constructing a metric space allows us to evaluate vectors for relative similarity based on quantitative measurements.  We can quantitatively evaluate documents for the degree to which they correspond to a topic, and which datasets are have clear exemplars of topics with high strength based on their similarity.

By multiplying the matrix $D$ of weights per dimension with the matrix $U$ of terms, and selecting the dimensions with the highest weight, we are able to plot vectors in a dimension that captures a majority of the variance for each term frequency vector. This dimension represents the degree to which an opcode relates to a higher level abstraction of program structure.

If we construct a metric space over $U$ obtained from SVD with a distance metric selected being Cosine similarity, then we have a basis for comparing the similarity of blocks by their terms.

Figures 4 and 5 show a metric space constructed from the highest weighted topics obtained from SVD.  Each dimension in the metric space represents one of the two highest weighted dimensions.  This gives us a spatial representation of the topic abstraction, and its relation to the documents.  We discuss the applications of this representation in Section 3.

\begin{figure}[t]
    \centering
    \hspace{-2cm}
    \includegraphics[width=1\textwidth]{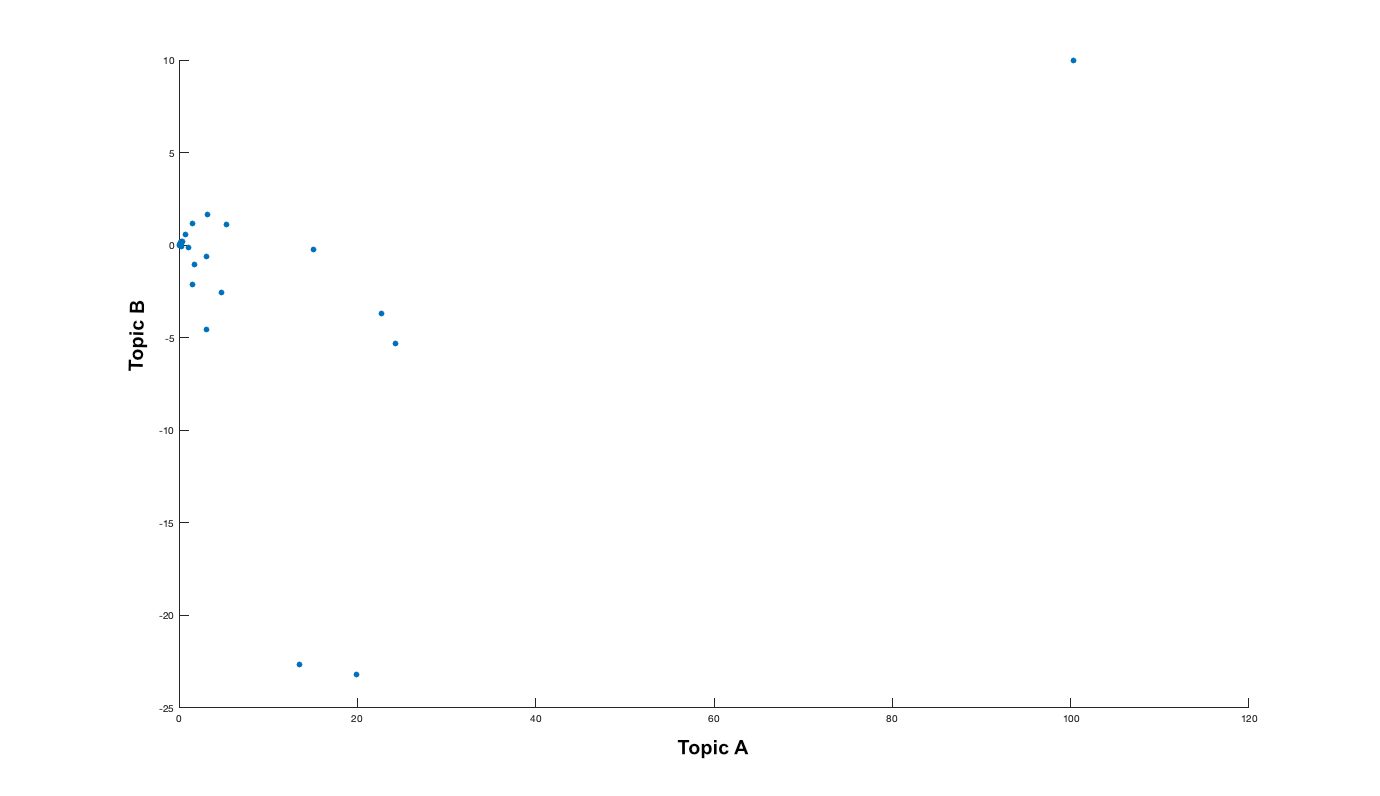}
    \caption{Metric space constructed from highest weighted topics.  Individual vectors correspond to term frequency vectors.  This shows the contribution of individual terms in the dictionary to the topics discovered.  The dimensionality of each vector is 32.  Using the topic projection we can evaluate each vector in a 2 dimensional Euclidean space.  Both topics have a high degree of correlation with data movement.}
\end{figure}

\begin{figure}[h]
    \hspace{-2cm}
    \includegraphics[width=1.2\textwidth]{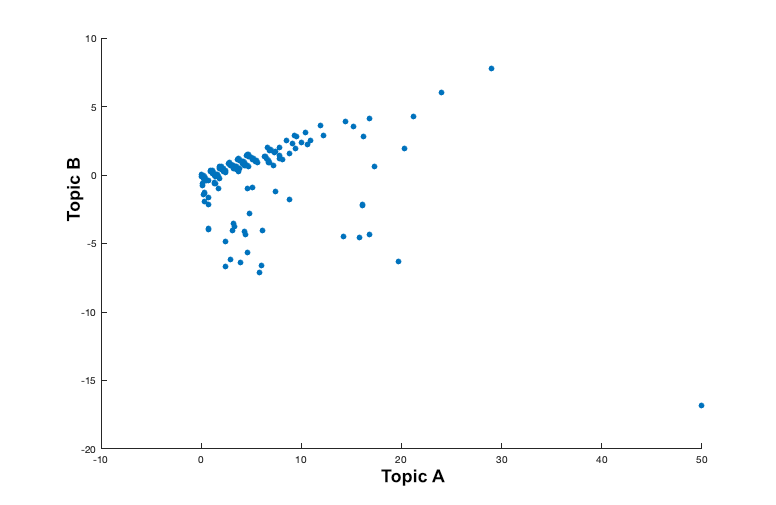}
    \caption{Metric space constructed from highest weighted topics for the same program and using the same topics.  This shows the complete representation of the program as a set of segment vectors.  Each vector represents a basic block segment, and can be viewed as a node in the program's control flow graph.  A spatial representation allows for an increase in resolution in comparison to analysis and hashing of whole document feature representations.  A single program is a collection of segments, and the program's impact on the topic is distributed across the set of segments.  We can see that the two topics differ in their representation of the program, as specific vectors have a larger magnitude in single dimensions.}
\end{figure}

\section{Results}
This section discusses the results of the experiment performed in section 2.

\subsection{Discussion}
The goal of this work is to provide an additional level of abstraction in order to determine the patterns generating the syntactic elements of malicious program binaries.  The features used for this are a vector representation using $tf-idf$ term frequency methods and the program's control flow adjacency matrix representation.  These features are extracted based upon segmentation of the program, which provides additional feature resolution.  The abstract representation is obtained from matrix decomposition using Singular Value Decomposition (SVD).  Metric spaces are defined for the program vectors as well as their topic abstractions in order to evaluate program features for similarity.  Once abstract topic representations are obtained from matrix decomposition, these topics are used to evaluate the original static artifacts of the program by projecting the vector for each program segment into the metric space defined by the topic.  In this section we explore a single program in metric spaces of topics selected from the component matrices.

Three resulting matrices are obtained from each matrix decomposition.  The matrix $U$ is an orthonormal basis for instructions in a segment which shows the relationship between a document and its abstraction. The matrix $D$ is a matrix of topic weights per dimension which shows the strength of the individual abstract topic representation.  The matrix $V$ is an orthonormal basis for programs which shows the degree to which a program is associated with an abstract representation or topic.  These components were evaluated individually using a spatial representation.

The goal of the matrix decomposition is to obtain a set of latent variables from each of the matrices.  These variables would represent a conceptual abstraction from the previous resolution of the data.

The square matrix $D$ obtained from performing SVD on an adjacency matrix representation of the control flow graph for a given program shows the abstract representations of the network.  The distribution of weights for each dimension is shown in Figure 3.  This figure shows that there are a large number, a number of unique topics approximately 30\% of the number of nodes in the network, each with a low weight (less than 10).  The space of nodes and the network were then reconstructed using the topics to create a scatter plot of the metric space obtained from decomposition of the control flow network \cite{hagberg2008exploring}.

One view of this representation is that a program is composed of a corpus of "documents", which are basic block segments.  Each basic block is a segment of the complete program.  The term frequencies in each vector represent the document term frequency of instructions in a block.

The decomposed matrix $U$ from the opcode frequency matrix is representative of how an opcode relates to a given abstraction.  The matrix $D$ gives the weight of each abstract representation.  The matrix $V$ represents how a basic block relates to a given abstraction.  This makes the assumption that term co-occurrence is significant.

Figure 4 shows a projection of the individual terms into the metric space defined by the highest weighted topics.  Individual vectors correspond to term frequency vectors selected from matrix $U$ using topics from matrix $D$. This shows the contribution of individual terms in the dictionary to the topics discovered. The dimensionality of each vector is 32. Using the topic projection we can evaluate each vector in a 2 dimensional Euclidean space.  If we view correlation as the vector cosine, we can see that both topics $A$ and $B$ have a low degree of correlation with a high number of terms and a very high correlation with a single term.  Both topics have a high degree of correlation with data movement.  This finding is validated by the distribution of term frequencies across the corpus.\cite{musgrave2020semantic}

Figure 5 shows a scatter plot of term frequency vectors.  Vectors are selected from the matrix $V$ using the topics from matrix $D$.  This projection shows a plot of the entire program represented as a set of vectors in the metric space.  This figure shows us how a program contributes to a given topic.  Each vector is plotted in a spatial representation, and corresponds directly to a basic block segment.  The segments were selected from a document corpus, which composes a program.  The program shown in this figure is a collection of vectors in the metric space.  Vectors in the space are plotted in the highest weighted dimensions obtained from SVD on each axis in two dimensions.  The dimensions were selected from the topic matrix, and used to construct the vector space.  This vector space provides a greater level of abstraction is suitable for further analysis into which documents are similar, and which documents are representative of specific concepts.  For example, the value of the cosine similarity in outliers has a value that is drastically different from that of a majority of the data points in the dataset.  This also shows a correlation that would be subject to regression analysis, as a linear trend is visible at several points in the space.  We hope to explore clustering in the topic space in future work.

In this figure we can see the contribution of individual vectors in the topic space from the vector magnitude.  It is important to note in this example that topics $A$ and $B$ differ in that the vector with the largest contribution to topic $A$ is not present in topic $B$.  This is an outlier, and would be a good candidate for a differentiating feature of this abstract topic representation.  This syntactic feature has a correspondence from the topic abstraction.  This represents an increase in the feature resolution that would not have been possible without the segmentation of the program document.

The adjacency matrix of a program's control flow decomposes into matrices of weights and space for nodes in control flow network.  As previously discussed, the abstraction shows a significantly large distribution of relatively small weights.  This implies that we are given a large number of very small features to detect, and their abstractions are not correlated.  This can be seen again when plotting node and network level embeddings of the topics, which are close to their respective axes and do not show a strong correlation via cosine similarity.  Program control flow networks appear to have properties that do not easily allow for decomposition, due to the low weighting of a large number of dimensional weights obtained in the square matrix after decomposition.  This indicates that there are not clear abstractions for this structure.

Figure 4 shows a metric space for each of the dimensions with the highest weights as obtained from SVD.  The vectors in this metric space are the right-singular vectors, which represent the degree to which a document corresponds to a more abstract representation, such as a topic as a collection of terms for natural language processing.  In this context, a "document" is a deterministic segment of a program, as a collection of operations.  This figure shows each of the program segments plotted in the dimensions that capture the highest variance.

We can see a clear cluster of similar segments of the program that have high to moderate magnitude in the y-dimension.  We can also see at least three outliers, two with low magnitude in each dimension, one with high magnitude in both dimensions.  While the magnitude of these vectors differs, they would be similar in terms of a cosine similarity metric.

\section{Conclusion}
In this study we have collected a dataset of term frequencies based on a dictionary.  We have reduced the dimensionality of the dataset by using a stemming approach from over 500 terms to 32 terms.  We have segmented the dataset into sequential blocks and an adjacency matrix of control flow representing the structure of sequences to provide an increase in feature resolution.  Segmentation was done for increased accuracy and a more fine grained resolution.  The control flow adjacency matrix was collected to evaluate the structural properties of the document.  When these matrices were decomposed to obtain an abstract representation of the structure, low weights of a large number of dimensions obtained in the square matrix after decomposition, indicating that there are not clear abstractions for this structure.  We have decomposed the datasets collected by using Singular Value Decomposition.  This was done to provide an increased level of abstraction to the term frequency representation.  This was measured by the strength of the topics in the matrix representing the topic weights.  The value in the topic matrix provides a measurement of the topic weight.  The weight of this topic is the strength of the abstraction discovered.  We have evaluated the data using a spatial representation by constructing a metric space for the highest weighted abstract representations.  We have evaluated the similarity of vectors in the metric space by measuring their cosine similarity.  By providing structure, abstraction, and accuracy, we can analyze structural patterns, find generative processes for syntactic patterns through abstraction, and show the correlation between representations.

\subsection{Acknowledgements}
This research was supported in part by Air Force Research Lab grant \#FA8650 to the University of Cincinnati.

Conflicts of Interest: The authors declare no conflicts of interest.
%
%
%
%

\bibliographystyle{splncs04}
\bibliography{references}

\begin{thebibliography}{10}
\providecommand{\url}[1]{\texttt{#1}}
\providecommand{\urlprefix}{URL }
\providecommand{\doi}[1]{https://doi.org/#1}

\bibitem{opcoderef}
X86 opcode and instruction reference,
  \url{http://ref.x86asm.net/#HTML-Editions}

\bibitem{axler1997linear}
Axler, S.: Linear algebra done right. Springer Science \& Business Media (1997)

\bibitem{chandrasekaran2020context}
Chandrasekaran, M., Ralescu, A., Kapp, D., Kebede, T.M.: Context for api calls
  in malware vs benign programs. In: International Conference on Modelling and
  Development of Intelligent Systems. pp. 222--234. Springer (2020)

\bibitem{djaneye2019static}
Djaneye-Boundjou, O., Messay-Kebede, T., Kapp, D., Greer, J., Ralescu, A.:
  Static analysis through topic modeling and its application to malware
  programs classification. In: 2019 IEEE National Aerospace and Electronics
  Conference (NAECON). pp. 226--231. IEEE (2019)

\bibitem{dumais2004latent}
Dumais, S.T.: Latent semantic analysis. Annual Review of Information Science
  and Technology (ARIST)  \textbf{38},  189--230 (2004)

\bibitem{griffiths2007topics}
Griffiths, T.L., Steyvers, M., Tenenbaum, J.B.: Topics in semantic
  representation. Psychological review  \textbf{114}(2), ~211 (2007)

\bibitem{hagberg2008exploring}
Hagberg, A., Swart, P., S~Chult, D.: Exploring network structure, dynamics, and
  function using networkx. Tech. rep., Los Alamos National Lab.(LANL), Los
  Alamos, NM (United States) (2008)

\bibitem{hennessy2011computer}
Hennessy, J.L., Patterson, D.A.: Computer architecture: a quantitative
  approach. Elsevier (2011)

\bibitem{hopcroft2001introduction}
Hopcroft, J.E., Motwani, R., Ullman, J.D.: Introduction to automata theory,
  languages, and computation. Acm Sigact News  \textbf{32}(1),  60--65 (2001)

\bibitem{kebede2017classification}
Kebede, T.M., Djaneye-Boundjou, O., Narayanan, B.N., Ralescu, A., Kapp, D.:
  Classification of malware programs using autoencoders based deep learning
  architecture and its application to the microsoft malware classification
  challenge (big 2015) dataset. In: 2017 IEEE National Aerospace and
  Electronics Conference (NAECON). pp. 70--75. IEEE (2017)

\bibitem{manning1999foundations}
Manning, C., Schutze, H.: Foundations of statistical natural language
  processing. MIT press (1999)

\bibitem{musgrave2020semantic}
Musgrave, J., Purdy, C., Ralescu, A.L., Kapp, D., Kebede, T.: Semantic feature
  discovery of trojan malware using vector space kernels. In: 2020 IEEE 63rd
  International Midwest Symposium on Circuits and Systems (MWSCAS). pp.
  494--499. IEEE (2020)

\bibitem{nar2019analysis}
Nar, M., Kakisim, A.G., Yavuz, M.N., So{\u{g}}ukpinar, {\.I}.: Analysis and
  comparison of disassemblers for opcode based malware analysis. In: 2019 4th
  International Conference on Computer Science and Engineering (UBMK). pp.
  17--22. IEEE (2019)

\bibitem{rawashdeh2021single}
Rawashdeh, O., Ralescu, A., Kapp, D., Kebede, T.: Single property feature
  selection applied to malware detection. In: NAECON 2021-IEEE National
  Aerospace and Electronics Conference. pp. 98--105. IEEE (2021)

\bibitem{sebesta1999programming}
Sebesta, R.W., Mon, T., day Mon, R., Class, L., Fri, N.: Programming languages
  (1999)

\bibitem{shafiq2009pe}
Shafiq, M.Z., Tabish, S.M., Mirza, F., Farooq, M.: Pe-miner: Mining structural
  information to detect malicious executables in realtime. In: International
  Workshop on Recent Advances in Intrusion Detection. pp. 121--141. Springer
  (2009)

\bibitem{shawe2004kernel}
Shawe-Taylor, J., Cristianini, N., et~al.: Kernel methods for pattern analysis.
  Cambridge university press (2004)

\bibitem{siddiqui2008detecting}
Siddiqui, M., Wang, M.C., Lee, J.: Detecting trojans using data mining
  techniques. In: International Multi Topic Conference. pp. 400--411. Springer
  (2008)

\bibitem{souri2018state}
Souri, A., Hosseini, R.: A state-of-the-art survey of malware detection
  approaches using data mining techniques. Human-centric Computing and
  Information Sciences  \textbf{8}(1), ~3 (2018)

\bibitem{steyvers2005large}
Steyvers, M., Tenenbaum, J.B.: The large-scale structure of semantic networks:
  Statistical analyses and a model of semantic growth. Cognitive science
  \textbf{29}(1),  41--78 (2005)

\bibitem{turi1997towards}
Turi, D., Plotkin, G.: Towards a mathematical operational semantics. In:
  Proceedings of Twelfth Annual IEEE Symposium on Logic in Computer Science.
  pp. 280--291. IEEE (1997)

\bibitem{witten1999weka}
Witten, I.H., Frank, E., Trigg, L.E., Hall, M.A., Holmes, G., Cunningham, S.J.:
  Weka: Practical machine learning tools and techniques with java
  implementations  (1999)

\end{thebibliography}
\end{document}